\documentclass{aa}  

\usepackage{graphicx}
\usepackage{txfonts}
\usepackage[hidelinks,colorlinks=true,linkcolor=blue,citecolor=blue]{hyperref}
\usepackage[dvipsnames]{xcolor}
\usepackage{multicol}
\usepackage{multirow}
\usepackage{makecell}
\usepackage{cancel}

\usepackage{float}
\usepackage{latexsym}
\usepackage{amssymb}
\usepackage{amsmath}
\usepackage{amsfonts}
\usepackage{color}
\usepackage{cool}
\usepackage{subfig}
\usepackage{dcolumn}
\usepackage{soul}
\usepackage[utf8]{inputenc}
\usepackage[english]{babel}
\usepackage{import}
\usepackage{tabularx}
\usepackage{verbatim}
\usepackage{natbib}
\usepackage{tabu}
\usepackage{bm}

\begin{document}

   \title{Probing the statistical isotropy of the universe with Planck data of the cosmic microwave background}

   \author{C. E. Kester
          \inst{1}
          \and
           A. Bernui\inst{2}
           \and
           W. S. Hipólito-Ricaldi\inst{1,3} \thanks{e-mail: wiliam.ricaldi@ufes.br
           }
}
   \institute{Universidade Federal do Esp\'{\i}rito Santo, Centro Universitário Norte do Espírito Santo, Departamento de Ci\^encias Naturais, Rodovia BR 101 Norte,  km. 60, S\~ao Mateus, ES,   29932-540, Brasil.
            \and
      Observat\'orio Nacional, Rua General Jos\'e Cristino 77, 
	S\~ao Crist\'ov\~ao,  Rio de Janeiro, RJ,  20921-400,  Brasil.
         \and 
            Núcleo de Astrofísica e Cosmologia (Cosmo-Ufes),
Universidade Federal do Espírito Santo, Vitória, ES, 29075-910, Brasil.
             }

   \date{Received xxx; accepted xxx}

   \abstract
  {}
  {
We study the angular distribution of temperature fluctuations in the cosmic microwave background (CMB) to probe the statistical isotropy of the universe by using precise full-sky CMB data with a model-independent approach. 
  }
  {
We investigated the temperature-temperature angular correlations in the four Planck foreground-cleaned CMB maps that were released recently. We performed a directional analysis on the CMB sphere to search directions in which the temperature-temperature angular correlations are extreme.
  }
{
Our analyses confirm a preferred axis in the CMB sphere, pointing in the direction $(l,b) \simeq (260^{\circ}, 130^{\circ})$, at the $98\% -99\%$ confidence level. In this direction, the CMB angular correlations exceed the antipodal direction most strongly. 
This preferred direction is unexpected in the $\Lambda$CDM cosmological model and represents a significant deviation from results obtained by applying the same procedure to simulated statistically isotropic CMB maps. 
This result confirms the north-south asymmetry in the most recent Planck data. This phenomenon is one of the previously reported  CMB anomalies.
}
{
We performed a robust detection of the north-south asymmetry in the temperature-temperature angular correlations, with a slightly different statistical significance, in the four Planck foreground-cleaned CMB maps. 
Moreover, we performed consistency tests by adding foreground and noise, both Planck data products, to the CMB map we studied, and we also investigated and discarded possible bias in our method. 
After these detailed analyses, we conclude that the north-south asymmetry phenomenon is present with a high 
statistical significance in the Planck CMB maps we studied. This result confirms previous reports in the literature in the past 20 years.
}

  

   \maketitle
%
\section{Introduction}
The angular distribution of temperature fluctuations in the cosmic microwave background (CMB) is a valuable cosmological probe for understanding primordial and late processes in the universe. Moreover, these data can also probe the statistical isotropy (SI) hypothesis, which is a fundamental attribute of the concordance cosmological model ~\citep{Jarosik2011,Planck18}. 
   
During the past 20 years, diverse analyses of the CMB data at large angular scales, both from the 
Wilkinson Microwave Anisotropy Probe 
(WMAP) and Planck satellites, have provided evidence of significant deviations at various confidence levels of the SI hypothesis. These features are known in the literature as  CMB anomalies~\citep{Pla18-ISO}. 
These intriguing features are important for understanding the global properties of the universe, such as its spatial topological shape. It is therefore valuable to determine whether the reported  statistical anisotropy is cosmological in origin, and, if so, 
extract information that may be helpful for identifying its causes. 
To avoid possible bias related to a specific analysed dataset or a particular estimator, 
it is worthwhile that different studies from diverse research groups outside the Planck Collaboration team can perform complementary analyses~\citep{Pla18-ISO}. 


Since the early release of WMAP data, many analyses have reported unexpected features in the angular distribution of the CMB, mainly at large scales. 
Among these are the low-quadrupole amplitude of the CMB angular power 
spectrum~\citep{Bennett2003,Jarosik2011}, the alignment of some low-order multipoles~\citep{deOliveira-Costa2003,Bielewicz2004,Schwarz2004,Bielewicz2005,Land2005,Abramo2009,Polastri2015}, 
the lack of large-angle temperature-temperature (TT) correlations~\citep{Schwarz2004,Copi2006,Bernui06,Copi2009,Schwarz2016}, 
the parity asymmetry \citep{Kim2010,Gruppuso2011,Aluri2012,Hansen2012,Zhao2013,Aluri2017}, the cold spot \citep{Vielva2003,Cruz2005,Bernui09,Gurzadyan2014}, 
and the north-south asymmetry \citep{Eriksen2004a,Hansen2004a,Eriksen2004b,Hansen2004b,Hoftuft2009,Paci2010,Pietrobon2010,Rath13,Bernui2014,Gimeno2023}. 
Recently, it was also reported that the CMB lensing amplitude, $A_L$, has a value higher than expected in the $\Lambda$CDM 
model~\citep{Calabrese2008,Planck18,Mokeddem23}. 

Some methods that investigated SI features in the CMB data used cosmology-dependent approaches. 
In this way, the obtained results are of restricted validity because the outcomes depend on the hypotheses that were assumed. 
In this scenario, the conclusions from model-independent analyses are 
preferable because they are universally valid. 
For this reason, we adopted a model-independent approach based on the two-point angular correlation function (2PACF) in spherical caps. This approach 
allowed us to perform a directional analysis of the TT correlations over the CMB sphere, searching for directions with outlier properties. 
Through this directional analysis, we measured possible deviations from the SI hypothesis by quantifying the statistical significance of our findings by comparison with the outcomes of a similar directional procedure applied to a large number of SI CMB maps~\citep{Bernui07}. 
We applied this method to the latest Planck CMB temperature-fluctuations maps, that is, the four foreground-cleaned Planck CMB maps, and we also considered in the analysis the effects caused by residual foregrounds and instrumental noise because they might affect our results. 
In particular, our directional method is optimal for detecting preferred directions (i.e., asymmetries) on the CMB sky.

Our directional approach also allowed us to study the CMB at different large scales (different $\gamma_0$ radius, as we discuss below). This technique can provide information about the cause(s) of the anomalous behavior. 
Applications of this method include, for instance, the study of the 
Galactic masks used for 
analyses of WMAP data, concluding that they weakly affect the direction and statistical significance of the 
north-south asymmetry ~\citep{BMRT07}. 
In addition, we also investigated that the cold-spot anomaly does not cause the north-south asymmetry~\citep{Bernui09}
(for other interesting results using this directional analysis approach, see, e.g.,~\citet{Marques2018} for the analysis of the Planck convergence map, 
or~\citet{BFW2008} for the study of the large-angle distribution of the GRB). 
Additionally, because several reported CMB anomalies are related to angular correlations or to preferred directions, an approach that combines the two features is highly welcome. Our method is based on 2PACF and shows a preferred axis on the sky, indicating a possible connection with the CMB quadrupole-octopole alignment (see the discussion in~\cite{B-HR} 
for a universe with preferred axis, sourced by a primordial magnetic field, in the analysis of WMAP data, and in our 
Appendix~\ref{appendix} for the general case considering the analysis of the latest Planck data).


On the other hand, even though the north-south asymmetry phenomenon has been reported since 2004~\citep{Eriksen2004a,Hansen2004a}, the study of precise CMB data sets from WMAP and Planck 
using diverse procedures and algorithms, an exhaustive reanalysis of the latest and most precise data (including the use of Planck CMB products) allows us to capture possible systematics or data-processing effects on the north-south phenomenon. It also allows a reevaluation of its statistical significance. 
Our main contributions in studying this phenomenon are: a)  In a model-independent way, the latest Planck CMB maps show an SI violation and north-south asymmetry. The results for both the direction and the statistical significance are robust. We found this 
in all the foreground-cleaned maps and through consistency tests; b) we show that SI violation and north-south asymmetry in the latest Planck CMB maps are not correlated  with instrumental noise, Galactic foregrounds, or known systematics; c) with our directional analysis (which is described in detail in section~\ref{sec3}) 
we found an anisotropic behavior of the angular correlations, 
revealing that a large part of the sky has small large-angle correlations. 
This is probably related to another CMB anomaly: the lack of power of the CMB large-angle correlations~\citep{Copi2006}; d) we found that the asymmetry in the angular correlations appears at the largest scales. We discuss (see Appendix~\ref{appendix}) the possible relation of this to the quadrupole-octopole multipole alignment.

Differently from studies that assumed a fiducial model to study the CMB north-south asymmetry with the angular power spectrum 
$\{ C_{\ell} \}$ 
(see, e.g.,~\citet{Eriksen2004a,Hansen2004a,Rath13,Yeung2022}), or studies that assumed a cosmological model with a dipolar modulation  a priori
(see, e.g.,~\citet{Rath2013b,Mukherjee2015,Planck16-geo,Gimeno2023}), 
our approach is a model-independent method that is based on a directional analysis using the 2PACF as estimator to probe cosmic isotropy. 
Consequently, all our findings here are also model independent.

The outline of this work is the following. 
In section~\ref{sec2} we briefly present the Planck CMB data that we investigated with our directional analysis procedures. 
In section~\ref{sec3} we describe the method 
of this SI scrutiny, and we also detail how the statistical significance of possible outlier directions can be quantified. 
In section~\ref{sec4} we present our results, discuss consistency tests to determine whether they are indeed robust, and in section~\ref{sec5} we formulate our conclusions and final remarks.


\section{Planck CMB data}\label{sec2}

In 2020, the Planck Collaboration released the fourth set of CMB products (PR4) derived from the Planck mission\footnote{Based on observations obtained with Planck (\url{http://www.esa.int/Planck}), 
an ESA science mission with instruments and contributions directly funded by ESA Member States, NASA, and Canada.}. 

These data products include the four Planck foreground-cleaned CMB temperature maps~\citep{Planck18-IV}, which are called Spectral Matching Independent Component Analysis (SMICA), Needlet Internal Linear Combination (NILC), Spectral Estimation Via Expectation Maximization (SEVEM), 
and  Bayesian parametric method Commander (\texttt{Commander}). 
These high-resolution maps have $N_{\text{side}} = 2048$ and an effective beam: 
full width at half maximum (FWHM) 
\texttt{fwhm} = $5^{\prime}$. 
These CMB maps result from diverse data analyses that were used to remove known foregrounds, in which each component-separation algorithm processed the multi-contaminated sky regions differently, such as the Galactic plane region. As a result, they exhibit different foreground-cleaned regions for each procedure. 
These regions are defined outside the so-called component-separation 
confidence mask, which is released jointly with the associated CMB map,
corresponding to $f_{\text{sky}} \simeq 0.83, 0.77, 0.81$, and $0.87$, 
for the SMICA, NILC, SEVEM, and \texttt{Commander} maps, respectively \footnote{\url{https://pla.esac.esa.int/\#maps}}. 

Our main objective in this work is to analyze the angular correlations of these four foreground-cleaned Planck CMB maps in detail with our directional estimator~\citep{Bernui07}. 
An important point is the consistency tests, which we performed by applying our directional analysis to investigate these four CMB maps. For example, these maps might reveal different asymmetry directions, which might suggest a systematic effect and may not be the result of a cosmological origin.
After this, we repeat our analyses considering the addition of foregrounds and white noise, both Planck CMB data products, to the CMB foreground-cleaned Planck map in order to test the possibility 
that our results are sourced by residual contaminants.




\section{Method and implementation}\label{sec3}

Next, we describe the method we adopted to study the CMB angular correlations present in the foreground-cleaned Planck maps. 
These CMB maps have been investigated by the Planck collaboration with several consistency tests. They are therefore reliable for CMB analyses outside the region defined by the corresponding confidence mask. 

\subsection{Method}

Our interest here is to search for possible preferred directions of the large-scale TT angular correlations in the Planck CMB maps. 
To quantify these statistical features, we define the 2PACF as
\begin{equation}\label{2fcpa}
C(\gamma) \equiv \left \langle 
\Delta T(\theta,\phi)\,
\Delta T(\theta',\phi') \right \rangle, \end{equation}
where $\gamma$ is the angle between pixels in directions $\hat{n} =(\theta,\phi)$ and $\hat{n}' =(\theta',\phi')$, and it is given by 
\begin{equation}
\gamma \equiv \arccos{[\cos{\theta}\cos{\theta'} + \mathrm{sen}\theta \,\mathrm{sen}\theta'\cos{(\phi - \phi'})}] \,.
\end{equation}
The average $\langle \, \rangle$ is computed  over all pairs of pixels with an angular separation $\gamma$. 
The computation of this quantity over the entire celestial sphere gives no indication of possible outlier directions, although it shows an anomalous lack of strength at large-angle TT correlations (see, e.g.,~\cite{Copi2004}). 
Instead, we searched for preferred directions by computing the 2PACF of TT correlations in a set of spherical caps by scanning the full CMB sky.

We assumed that the CMB temperature-fluctuation sphere is pixelized. 
We defined a spherical cap with a radius equal to $\gamma_0$, centered on the $J$ pixel with angular coordinates $(\theta_J, \phi_J)$ as 
\begin{equation}
\Omega_{\gamma_0}^J \equiv \Omega(\theta_J, \phi_J; \gamma_0) \subset \mathcal{S}^2 \,,
\end{equation}
where $\mathcal{S}^2$ represents the celestial sphere. 
In each of these caps, the 2PACF was computed  by using 
\begin{equation}\label{2pacf}
C(\gamma)^J \equiv \left \langle 
\Delta T(\theta_J,\phi_J)\,
\Delta T(\theta'_J,\phi'_J) \right \rangle\ ,
\end{equation}
with $\gamma \in \langle\, 0, 2\gamma_0\, ]$~\footnote{The symbol $\langle \cdots,\cdots ]$ means that the interval is open by the left and closed by the right.}. We denote by $C_k^J \equiv C(\gamma_k)^J$ the 2PACF value for the angular distances inside the subinterval  $\gamma_k \in \gamma$, such that $\gamma_k \in \langle\, (k-1)\beta, k\beta\, ]$ for $k = 1,..., N_{\textrm{bins}}$, 
where $N_{\textrm{bins}} = 2\gamma_0 / \beta$ is the number of bins and $\beta$ is the bin length. 
   
The function $C = C(\gamma)$ measures the 2PACF in different directions because it encompasses the information along different spherical caps. 
The next step was to quantify the difference between angular correlations observed in different directions. 
In order to do this, we used the $\sigma$ map 
method~\citep{Bernui07,Bernui08}. 
The $\sigma$ map is a model-independent method for detecting 
directions in the CMB maps with different angular correlation intensities. The analysis made through the 2PACF on spherical caps that scan the celestial sphere and then compare the observational data results with the same procedure applied to synthetic SI maps. 
With this method, $\sigma$ maps are celestial maps based on real values, which are 
then converted into colored maps, containing a measure of the angular correlations from each region of the celestial sphere. 
To quantify the correlation intensity, it is possible to define a scalar function to associate a non-negative real value to each spherical cap with center in $(\theta_J,\phi_J)$, that is, 
\begin{equation}
\sigma_J: \Omega^J_{\gamma_0} \subset \mathcal{S}^2 \mapsto \mathbb{R}^+\ ,
\end{equation}
for $J = 1, ... , N_{\textrm{caps}}$, where $\sigma_J$ 
is defined by 
\begin{equation}\label{sigma}
\sigma_J^2 \equiv \frac{1}{N_\text{bins}}\sum_{k=1}^{N_{\mathrm{bins}}}(C_k^J)^2 \,,
\end{equation}
with $\sigma_J=\sigma_J(\theta_J,\phi_J)$. 
Then, the set of real positive numbers $\{ \sigma^2_J \}$ for all $J = 1,\cdots,N_\text{caps}$ is the  $\sigma$ map. 
To detect any possible asymmetry in any direction, 
we  decomposed the $\sigma$ map in   spherical harmonics,
\begin{equation}\label{sigma1}
\sigma^2(\theta,\phi) = \sum_{\ell m} a^{\sigma}_{\ell m} Y_{\ell m}(\theta,\phi)\,,
\end{equation}
and analyzed its power spectrum,
\begin{equation}\label{aps}
\text{S}_\ell = \frac{1}{2\ell + 1}\sum_{m = -\ell}^{\ell}|a^{\sigma}_{\ell m}|^2\ .
\end{equation}
The mean, the standard deviation, and the distribution of values $\{ \text{S}_\ell \}$ 
of the $\sigma$ map power spectra obtained from a directional analysis applied to a set of SI CMB maps, help us to quantify how frequent or rare the power spectra of the Planck $\sigma$ maps are (i.e., the $\sigma$ maps obtained by analyzing the four foreground-cleaned CMB maps). 
A high value for the dipole $\text{S}_1$, for instance, would indicate a preferred axis on the sky, that is, a possible violation of the SI of the universe.

\subsection{Angular correlation analyses}

To implement the method presented in the previous subsection and to obtain a quantitative measurement of possible directional differences in the angular correlation of a CMB map, we covered the sky with  $N_\text{caps}$ spherical regions and  computed the set of $\{\sigma_J, J = 1,\cdots, N_{\textrm{caps}}\}$ values, that is, its $\sigma$ map. 
We therefore followed the next steps: i) we downgraded the CMB maps and its respective masks, and ii) we generated the set of spherical caps with a radius $\gamma_0$ with CMB data from a given Planck masked map; iii) we then computed the TT 2PACF in all these spherical caps; and iv) we produced the Planck $\sigma$ map and obtained its angular power spectrum.
As discussed above, we analyzed the four Planck foreground-cleaned CMB temperature maps~\citep{Planck18}. 
These maps have a high resolution with $N_\text{side} = 2048$ and $50.331.648$ pixels, which corresponds  to a multipole  $\ell \sim 2500$. 
Because several anomalies were reported at large angular scales, that is, $\ell_{max} \le 30 $ (see, e.g.,~\cite{deOliveira-Costa2003, Eriksen2004a,Copi2006,Bernui07,Copi2009}), we are interested here in studying a possible violation of the SI hypothesis at these angular scales. 
To do this, it is necessary to downgrade each of the original Planck foreground-cleaned maps. 

We downgrade the four Planck maps from $N_\text{side} = 2048$ to $N_\text{side} = 32$, which corresponds to $\ell_{max} = 95$ and $12 288$ pixels. 
The same downgrade process was required for their corresponding masks. 
The original masks have pixels with 0 and 1 boolean values. 
During the downgrade process, several pixels are combined to form one single pixel. In this process, the numerical value in each new pixel is the mean value of all mixed pixels. In general, these values are not 0 nor 1. It is therefore necessary to define a threshold as a criterion for reconstructing the 0 or 1 values. We defined this criterion by considering that pixels with values $\geq 0.8$ were considered as 1, and 0 otherwise. At the end of the downgrade step, we obtained a CMB map and its corresponding mask in the correct resolution to continue our analyses. The next step was to define the spherical caps for which the 2PACF was computed.

We defined the number and size of the spherical caps with CMB temperature data to analyze their 2PACF. 
We chose $\gamma_0 = 60^{\circ}$, but to confirm the consistency of our results, we finally also considered other values for $\gamma_0$ (see Table \ref{table2}). 
Our directional analysis performs a uniform scan of the CMB sky meaning that the centers of the spherical caps must be uniformly distributed. 
A practical way to define these cap centers uniformly is to use a  Healpix pixelization with a suitable angular resolution (i.e., $N^{\text{caps}}_\text{side}$). 
We chose $N^{\text{caps}}_\text{side} = 4$, which means that we analyzed the 2PACF in $N_{\textrm{caps}} = 192$ spherical caps. 
Clearly, other values for $N^{\text{caps}}_\text{side}$ can be considered to test the robustness of our results. 
To summarize, given a CMB map, our method consists of 
selecting for scrutiny 192 spherical caps of $60^{\circ}$ each and computing the 2PACF to the data inside each cap 
by means of equation~(\ref{2pacf}) to obtain a set of 192\, 
2PACF: $\{ C(\gamma)^J \}$ for 
$J=1,\cdots,192$~\footnote{In practice, the interval for calculating the 2PACF
$\langle\, 0, 2\gamma_0 \,] = \langle\, 0, 120^{\circ}]$ does not work equally well for the 192 spherical caps: Because of the Galactic mask that was applied, some caps have deficit 
of pairs at large angles. 
To solve this problem, we verified that 
scales smaller than $\sim \!100^{\circ}$ are suitable. We 
therefore computed the 2PACF in the interval $\langle\, 0, 100^{\circ}]$ for all spherical caps.}.

Then we applied equation~(\ref{sigma}) to this set of functions to obtain a set of 192 real positive numbers $\{ \sigma_J^2 \}$, 
which is full of directional information regarding the strength of angular correlations of the CMB map we analyzed. 
In practice, this set of numbers, $\{ \sigma_J^2 \}$, was converted into a colored map with 192 pixels, that is, the $\sigma$ map. This map has an angular resolution $N^\text{caps}_\text{side} = 4$. 
The angular power spectrum of the $\sigma$ map was then calculated, as usual, with equations~(\ref{sigma1}) and~(\ref{aps}). 
The power spectrum of the $\sigma$ map was then compared to the mean spectra of the corresponding $\sigma$ maps we obtained, according to the same procedure, from a set of $1 000$ SI CMB maps. 
This comparison provided the statistical significance of our directional analysis. 
Our results are presented in the next section.


\section{Results of the directional analysis}\label{sec4}

In this section, we present the results of the directional analyses 
we applied to the recently released Planck CMB maps. 
According to the method described in the previous section, we first obtained the corresponding $\sigma$ map of the analysed Planck CMB map, calculated its angular power spectrum, and then evaluated its statistical significance by comparison with spectra obtained by performing the same procedure applied to simulated CMB maps, 
called Monte Carlo (MC) CMB maps. 
These MC maps were generated by assuming the SI hypothesis using the concordance $\Lambda$CDM model angular power spectrum as seed \citep{Planck18}. 
Before we applied the directional analysis to these MC CMB maps, we verified that they shared the same observational features as the Planck maps, such as angular resolution and Galactic mask. 
This analysis allowed us to quantify the significance level of our findings and to evaluate whether the SI hypothesis was satisfied in the investigated Planck CMB data. 

\subsection{ \texttt{Commander}-$\sigma$ map}

In this section, we present the directional analysis of the \texttt{Commander} CMB map, the $\sigma$ map we obtained, its angular power spectrum, and other related results. 

\begin{figure}
\includegraphics[width=\columnwidth]{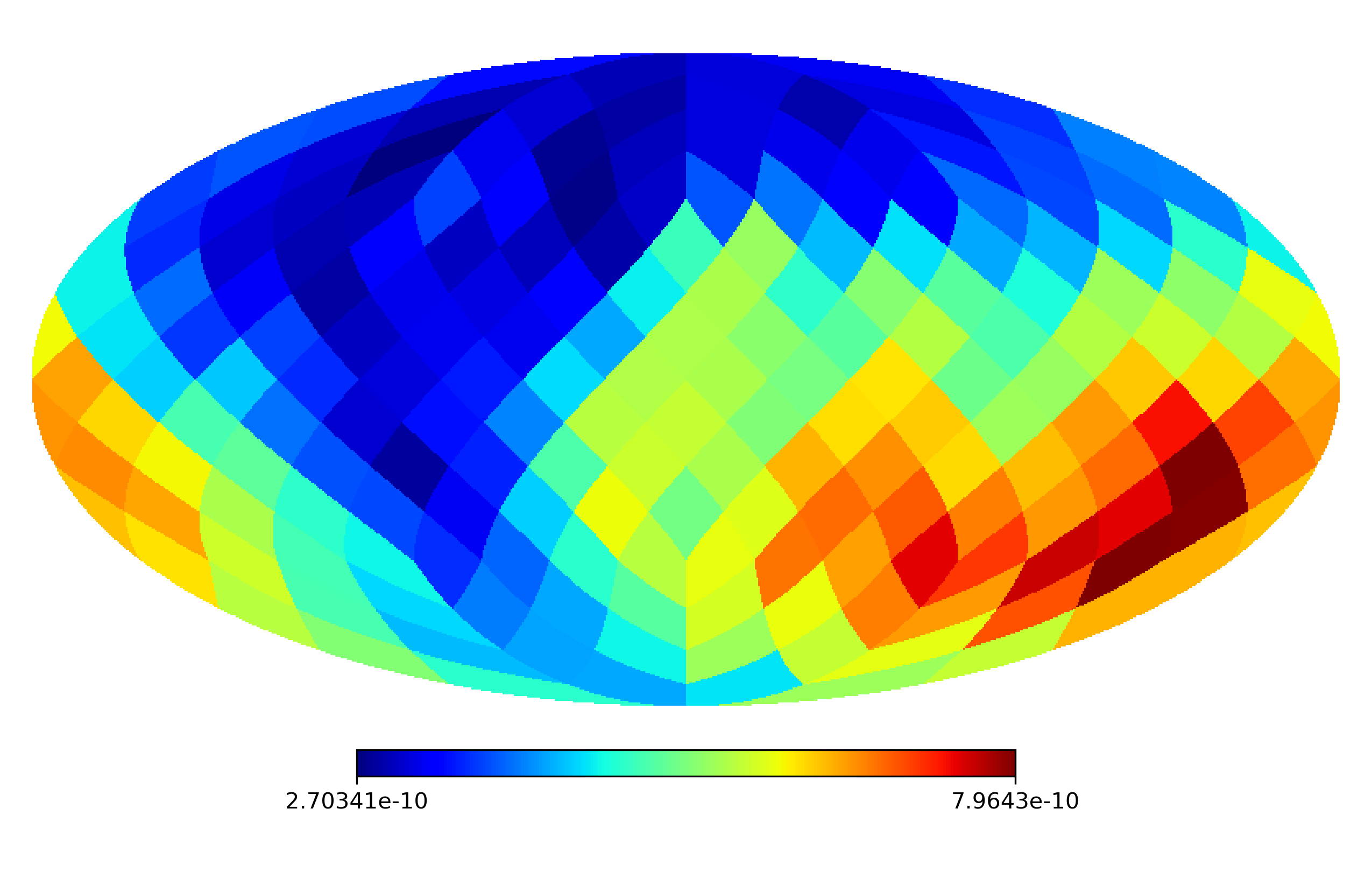}
\includegraphics[width=\columnwidth]{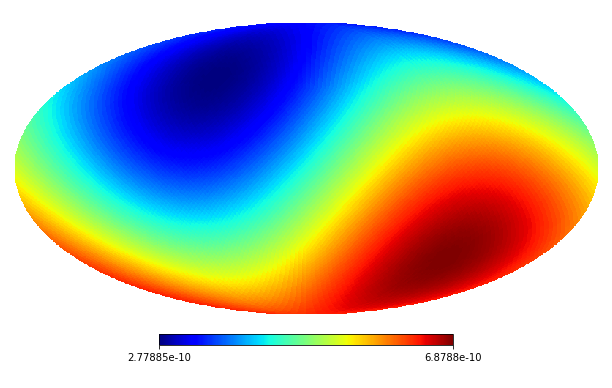}
\caption{
{\bf Top:} \texttt{Commander}-$\sigma$ map, 
which is the result of a directional analysis of the TT angular correlations in 192 spherical caps, with a radius $\gamma_0 = 60^{\circ}$, applied to the \texttt{Commander} Planck CMB map. Two extremes values in almost antipodal positions are visible.
{\bf Bottom:} Dipole component of the \texttt{Commander}-$\sigma$ map (top panel), which shows the dipole direction of the north-south asymmetry phenomenon: $(l,b) \simeq (253^{\circ}, 137^{\circ})$.
} 
\label{commander}
\end{figure}

\begin{figure} [h!]
 \includegraphics[width=\columnwidth]{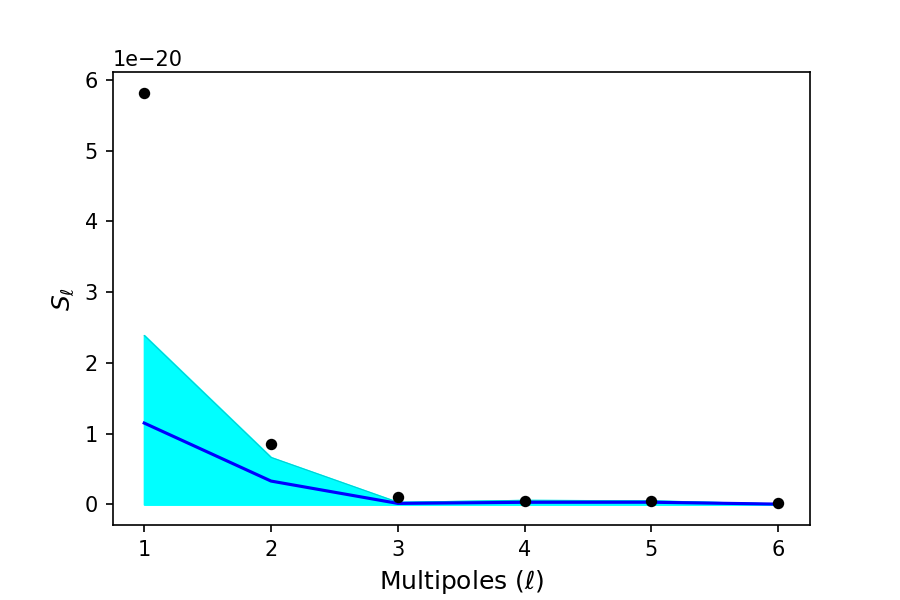}
 \caption{Angular power spectrum of the \texttt{Commander}-$\sigma$ map (dots), shown in Figure~\ref{commander}, where the continuous blue line corresponds to the mean angular power spectrum of the  \texttt{MC}-$\sigma$ map set, and the shaded region corresponds to 2 $\!\sigma$ CL.
}
 \label{powerspectrum}
\end{figure}

In the top panel of Figure~\ref{commander}, we show the result 
of our directional analysis applied to the \texttt{Commander} 
Planck map, that is, the \texttt{Commander}-$\sigma$ map. 
In the bottom panel, we display the dipole component 
of this map. 
The \texttt{Commander}-$\sigma$ map exhibits a dipolar red-blue region in the distribution of the angular correlations, which is strongly suggestive of a large-angle anisotropy. 
This can be better quantified in the \texttt{Commander}-$\sigma$ map power spectrum $S^{\texttt{Comm}}_\ell$, which is represented by dots in Figure~\ref{powerspectrum}.  By comparing $S^{\texttt{Comm}}_\ell$ with its corresponding  MC-$\sigma$ map power spectra, it is possible to evaluate whether they are consistent with each other.

After our directional analysis determined the dipolar feature in the analysed data, our objective was to quantify how common or rare this feature is in $\sigma$ maps produced from SI CMB maps. 
To do this, we applied the same method to a set of SI MC CMB maps and obtained a distribution of $\sigma$ map spectra $\{ \text{S}_{\ell}^{\text{MC}} \}$, for $\ell = 1,\cdots,5$. 
For our analyses with the \texttt{Commander} map, we generated $1 000$ MC maps and used the corresponding \texttt{Commander} fiducial mask to remove the Galactic region \footnote{\url{https://pla.esac.esa.int/\#maps}}. 
For each MC CMB map, we then computed the corresponding  $\sigma$ map (hereafter called MC-$\sigma$ map) and its angular power spectrum $\text{S}_\ell$. 
After the directional analysis of these $1000$ MC CMB maps, we calculated their corresponding angular power spectra. 
With this distribution  
in hand, we then calculated the mean and other statistical features. 
With this distribution, we calculated the statistical significance of $\text{S}_\ell^{\texttt{Comm}}$. 
The results are  presented in Figure~\ref{powerspectrum}. 
Moreover, the \texttt{Commander} CMB map was masked with its fiducial mask  to remove the  Galactic
region that might have residual foregrounds. We computed the \texttt{Commander}-$\sigma$ map  
and its power spectrum $\text{S}^{\texttt{Comm}}_\ell$, and finally, we compared it with their corresponding MC-$\sigma$ map results, as shown in Figure~\ref{powerspectrum}. 

Figure~\ref{powerspectrum} clearly shows that the $\text{S}^{\texttt{Comm}}_1$  dipolar component ($\ell = 1$) dominates the other higher-order multipoles ($\ell \ge 2$) and it is well outside of the 2$\sigma$ region, which means that its statistical significance is $> 95 \%$ CL. 
In fact, we calculate below that the dipole component $S^{\texttt{Comm}}_1$ is not consistent with the statistical isotropy hypothesis at $99 \%$ CL, as compared to MC-$\sigma$ maps obtained 
from a directional analysis applied to SI CMB maps.

On the other hand, the higher-multipole moments  $\text{S}^{\texttt{Comm}}_3$, $\text{S}^{\texttt{Comm}}_4$, $\text{S}^{\texttt{Comm}}_5$ 
are all inside the 2$\sigma$ region, 
except for the multipole moment $\text{S}^{\texttt{Comm}}_2$ ,
which is slightly above the 2$\sigma$ region. 
This could be due to an incomplete sky analysis or to the application of the confidence mask (see, e.g.,~\cite{BMRT07,Rassat2014}). 
Finally, we can conclude that the \texttt{Commander}-$\sigma$ map obtained from the \texttt{Commander} separation CMB map reveals a dipole moment $\text{S}^{\texttt{Comm}}_1$ that is larger than the most of dipole values in the set $\{ \text{S}^{\text{MC}}_1 \}$ that were obtained from the set of MC $\sigma$ maps (obtained from the directional analysis applied to the  MC CMB maps). 

\begin{figure}
\includegraphics[width=\columnwidth]{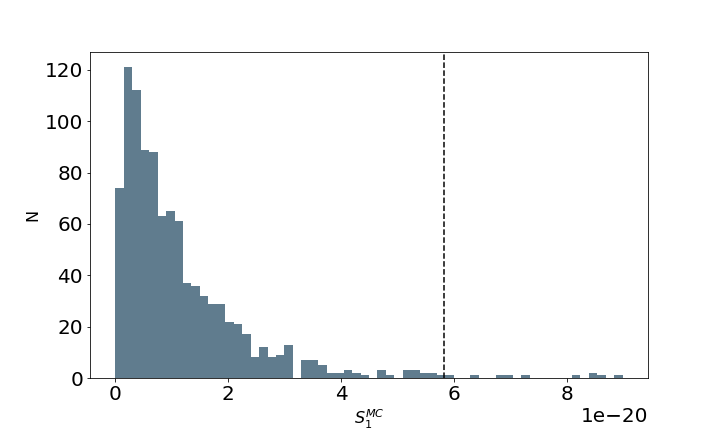}
\caption{Distribution of the $1000$ dipole moments from the MC-$\sigma$ maps, $\{ \text{S}^{\text{MC}}_1 \}$. The dipole moment of the \texttt{Commander}-$\sigma$ map, 
$\text{S}^{\texttt{Comm}}_1$, is included as a vertical dashed line. 
}
\label{probability}
\end{figure}

Additionally, it is possible to accomplish a probabilistic analysis involving the $\{ \text{S}^{\text{MC}}_1 \}$ dipole moment distribution for MC-$\sigma$ map power spectra  and  the \texttt{Commander}-$\sigma$ map power spectrum $S^{\texttt{Comm}}_1$. 
In other words, 
 we would like to determine how likely it is to find a value
$\text{S}^{\text{MC}}_1$ larger (or equal) than 
$S^{\texttt{Comm}}_1$. To know this, we constructed the $\text{S}^{\text{MC}}_1$ value probability distribution for all the  statistically isotropic MC-$\sigma$ maps  and  compared them with the value computed from the \texttt{Commander}-$\sigma$ map. 
The results are presented in Figure~\ref{probability}, where the $\text{S}_1$ value probability distribution is represented by 
dark blue bars  and  $\text{S}^\texttt{Comm}_1$ by a dashed line. 
The frequency with which $\text{S}_1$ is higher or equal to $\text{S}^{\texttt{Comm}}_1$ is small enough to reinforce our previous findings: one anomalously high  value for $\text{S}^\texttt{Comm}_1$. 
This was confirmed when we computed the  probability that $\text{S}^\texttt{Comm}_1 > \text{S}^{\text{MC}}_1$, which gives 
$\sim \!1.0 \%$ with respect to the SI case, as shown in Figure~\ref{probability} 
(see also Table~\ref{table1}).  

\begin{figure} [h!]
 \includegraphics[width=\columnwidth]{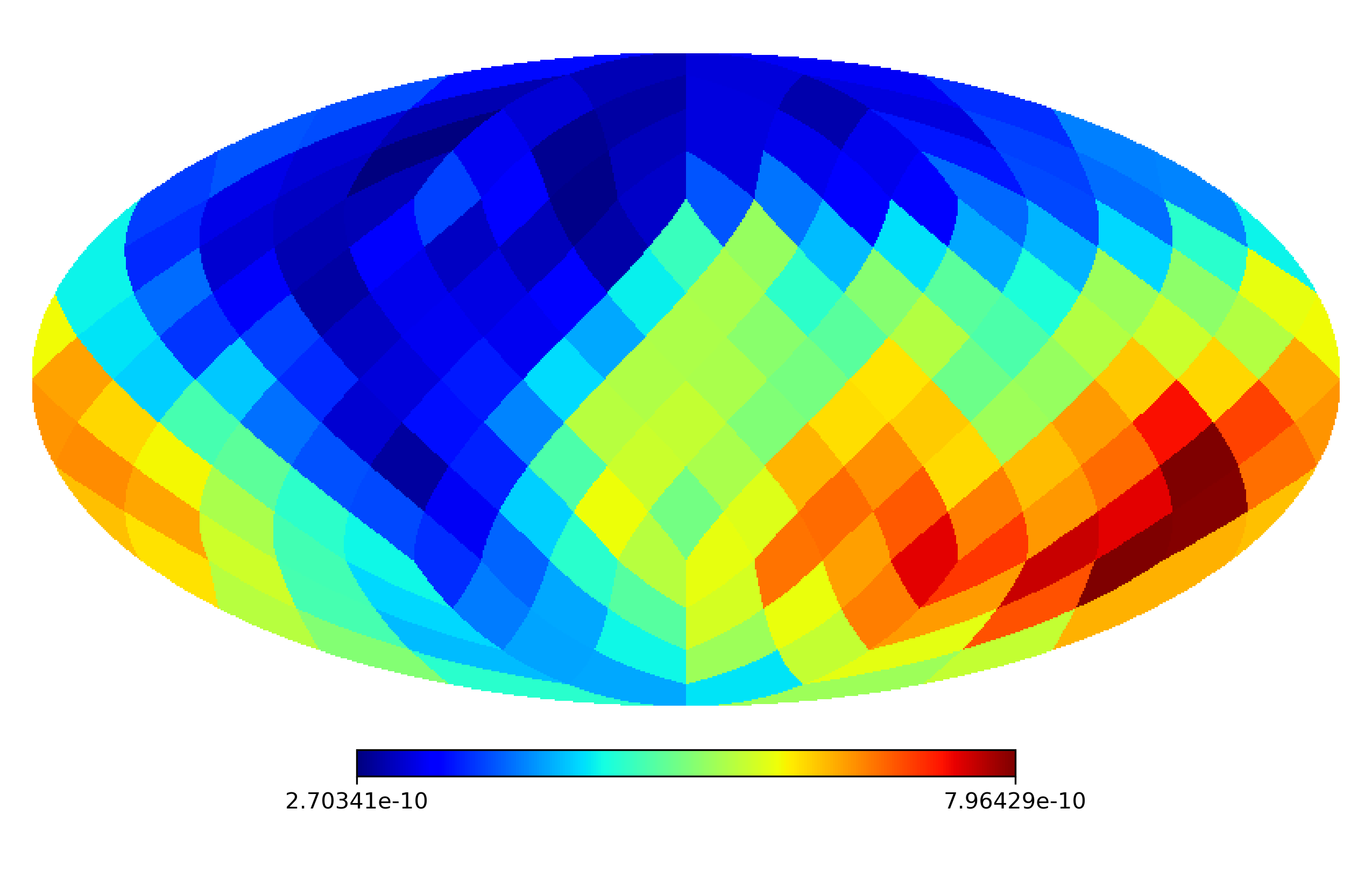} \\
 \includegraphics[width=\columnwidth]{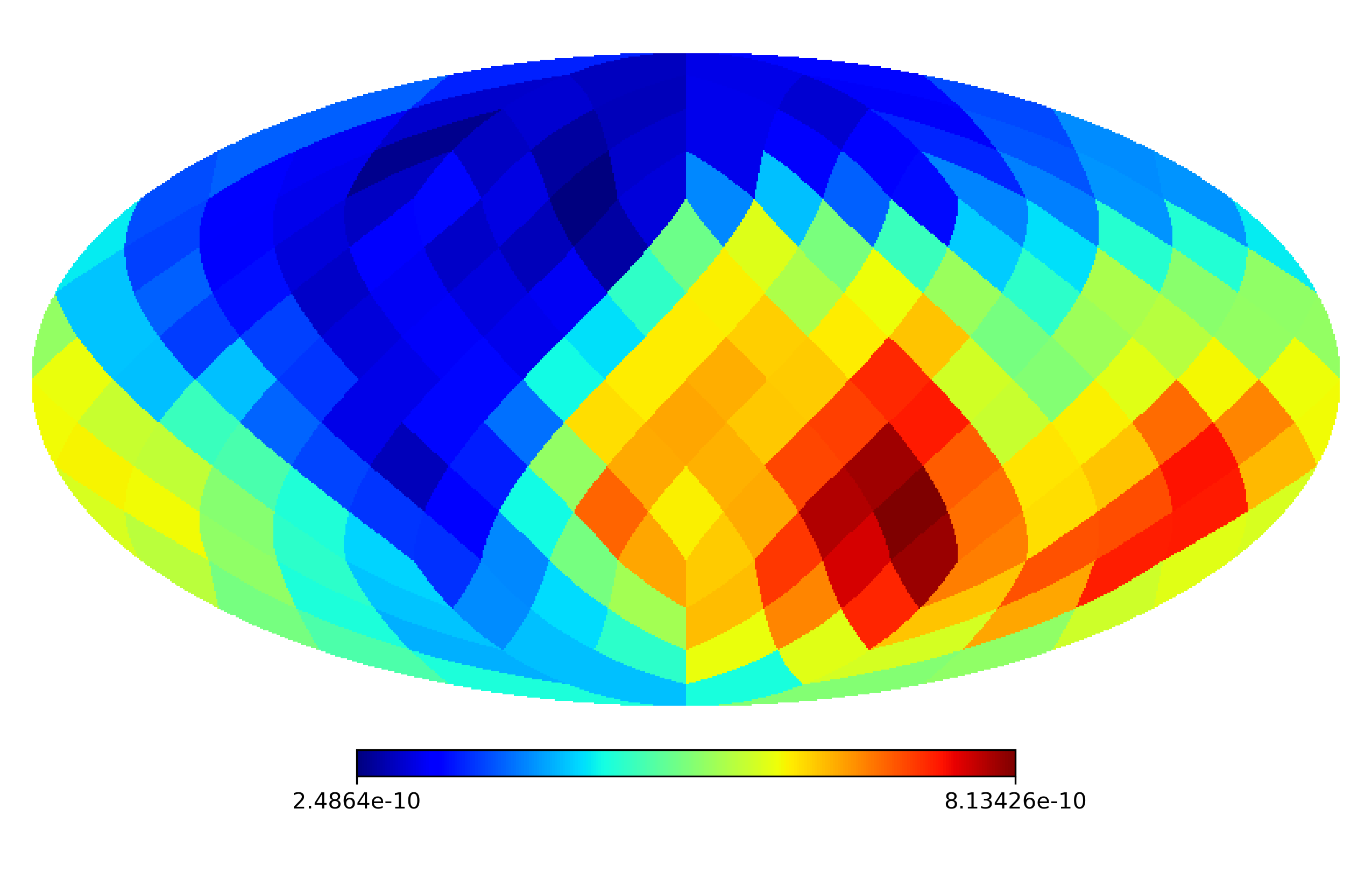} \\
 \includegraphics[width=\columnwidth]{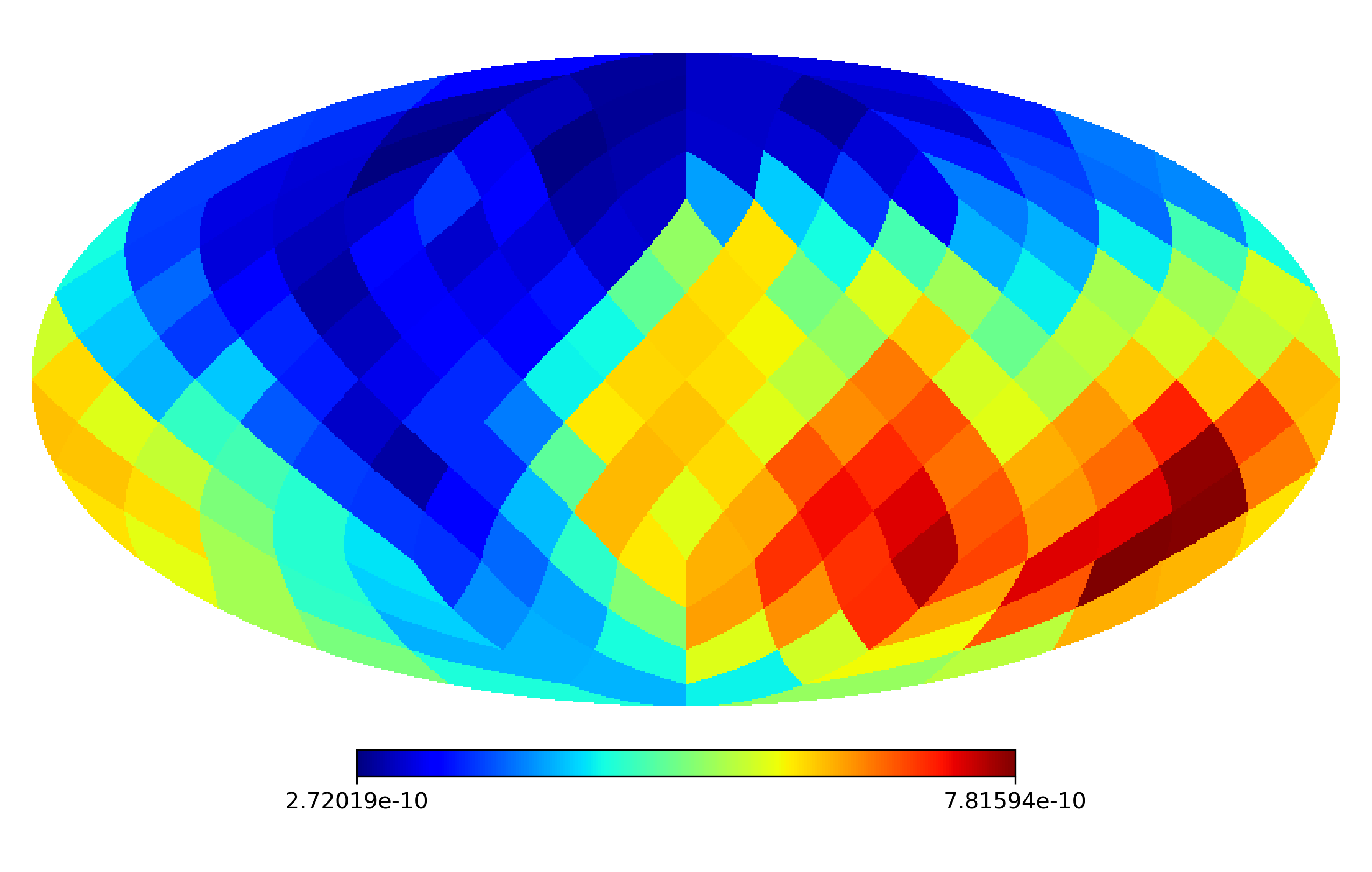}
  \caption{From top to bottom: 
NILC-$\sigma$ map, SEVEM-$\sigma$ map, and SMICA-$\sigma$ map. 
Observe that the high intensity regions (red pixels), corresponding to the caps with the largest TT angular correlations, are almost the same in all three maps. 
This interesting feature is confirmed with the 
measured dipole direction of these maps, quantities displayed in Table~\ref{table2}.}
 \label{separations}
\end{figure}

The $\sigma$ map captures the intensity of the angular correlations in all directions on the sky. 
By analyzing these maps, we can obtain insights about directions with stronger or weaker TT correlations. It is possible to detect any  anomalous difference in the TT correlations of two different directions in this way. 
Figure~\ref{commander} shows two almost antipode red-blue regions in the TT angular correlation distribution. 
It is then possible to evaluate the directions on the sky in which the correlation difference is maximum by searching the $\sigma$ map pixel with the highest and lowest intensities and finding their angular coordinates. Our findings show that the maximum and minimum correlations are always approximately antipodes. In Table~\ref{table1} we present  the angular coordinates for the direction along which the maximum correlation difference was found for the \texttt{Commander} case.  Furthermore, the same  features appear in the other three Planck separations NILC, SEVEM, and SMICA, as observed in Figure \ref{separations} and Table~\ref{table2}. This clearly shows that our results are quite robust.

Although we presented in this section the analyses considering only the \texttt{Commander} foreground-cleaned CMB map, in the  robustness studies discussed in the next subsection, we study the other three foreground-cleaned maps, NILC, SEVEM, and SMICA. 
We are interested in large-angle correlations. All the CMB map data, their corresponding masks, and  the MC maps were therefore downgraded to $N_\text{side} = 32$. 
This allowed us to concentrate our study of the SI on CMB maps in the angular scales $\ell = 2 - 95$.
\vspace{0.3cm}
\begin{table} [h!]
\small
\begin{tabular}{lcccc}
\hline
$\gamma_0 \,(^\circ$) &  Probability (\%) & Dipole $\times 10^{-20}$  & $b = \theta \,(^\circ)$ &   $l = \phi \,(^\circ)$ \\
\hline
 60  & 1.0 & 5.8 & 136.6 & 252.9 \\
\hline
\end{tabular}
\vspace{.1cm}
\caption{ \texttt{Commander}-$\sigma$ map analysis results for the 192 spherical caps of radius $ \gamma_0= 60^{\circ}◦$.}   
\label{table1}
 \end{table} 

\subsection{Robustness tests}

In this section, we present the results of several robustness tests, divided into two groups to support our previous findings, that is, that the asymmetry phenomenon manifests itself as a large dipole component of the \texttt{Commander}-$\sigma$ map, and its statistical significance. 
The first set of tests evaluated the possibility that our results might originate in  a systematic effect, such as residual foregrounds coming from the foreground-cleaning processes, large-angle correlations induced by the fiducial mask applied for each CMB map in the analysis~\citep{BMRT07}, or some residual noise present in the CMB maps. 
The second group of tests investigated the robustness by varying the parameters of the $\sigma$ map analyses, for example, the size of 
spherical caps used in the analyses, $\gamma_0$. 

\begin{center}
\begin{table*}
\centering
\begin{tabular}{lcccccc}
\hline
Data & $N_{\text{maps}}$ & $\gamma_0 \,(^\circ$) & 
$\%$ & Dipole $\times 10^{-20}$ & $b = \theta \,(^\circ)$ & 
$l = \phi \,(^\circ)$ \\
\hline
\texttt{Commander} & 1000 & 90 & 0.6
& 2.2 & 131.8 & 260.2\\
\texttt{Commander} & 1000 & 70 & 1.1
& 4.4 & 131.8 & 260.2\\
\texttt{Commander} & 1000 & 60 & 1.0 
& 5.8 & 136.6 & 252.9\\
\texttt{Commander} & 1000 & 45 & 0.7
& 12.2 &  133.4 & 262.7\\
NILC & 1000 & 60 & 2.2 
& 5.8 & 136.6 & 252.9\\
SMICA & 1000 & 60 & 1.4
& 6.1 & 135.0 & 265.5\\
SEVEM & 1000 & 60 & 1.8
& 6.9 & 133.4 & 271.5\\

\texttt{Commander} + Foregrounds & 1000 & 60 & 3.9
& 8.0 & 124.2 & 278.4\\
\texttt{Commander} + Noise & 1000 & 60 & 0.8
& 8.5 & 131.8 & 260.2\\
\texttt{Commander} + Noise & 1000 & 45 & 1.1
& 12.1 & 131.8 & 262.9 \\

\hline
\end{tabular}
\vspace{-.1cm}
\caption{
Directional analysis results. 
We show the direction of maximum values of the dipole component of the corresponding $\sigma$ map in each case.
The fourth column provides the percentage of $\sigma$ map dipoles obtained from SI MC CMB maps, greater than $\text{S}^{\texttt{Planck}}_1$, for each foreground-cleaned CMB map.
We perform these analyses for the four foreground-cleaned Planck maps, for various spherical caps sizes, and also considering the addition of noise and foregrounds to the original CMB map. 
For the sake of comparison, 
we included the results from the \texttt{Commander} map shown in Table~\ref{table1}. 
}\label{table2}
 \end{table*}
\end{center}

\begin{figure}
 \includegraphics[width=\columnwidth]{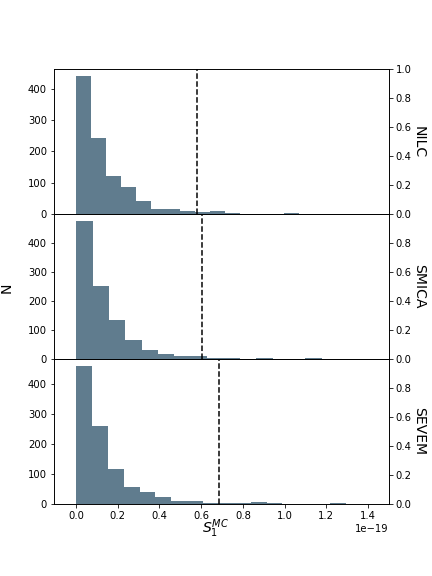} 
\caption{
Distribution of the dipole moments from three sets of $1000$ of MC-$\sigma$ maps (i.e., $1000$ values each for NILC, SMICA, and SEVEM), 
$\{ \text{S}^{\text{MC}}_1 \}$. 
We also plot together three vertical lines that correspond to the dipole moments of the NILC-$\sigma$ map, SMICA-$\sigma$ map, and SEVEM-$\sigma$ map.}
\label{3dist-dipolos}
\end{figure}

\begin{figure} [h!]
 \includegraphics[width=\columnwidth]{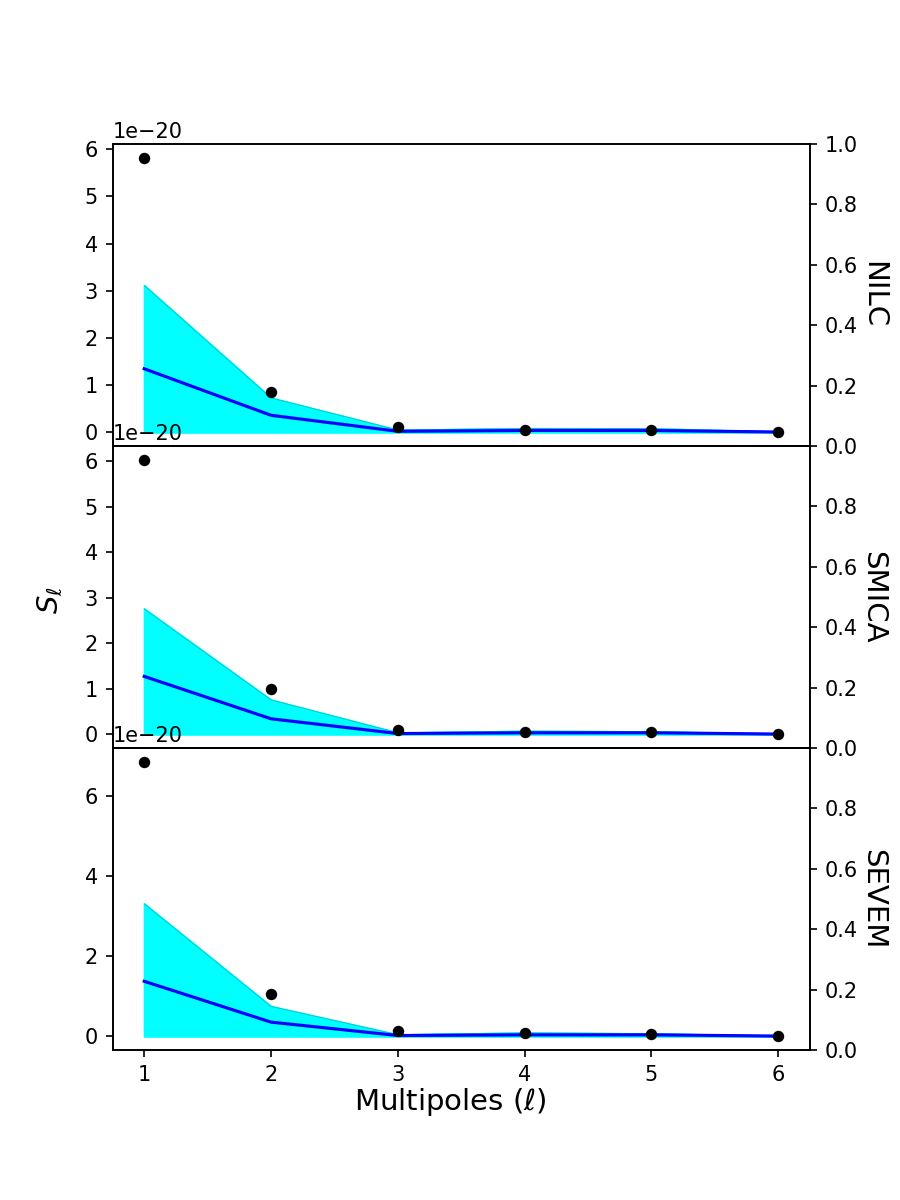}
 \caption{Angular power spectra of the NILC-$\sigma$ map, 
SMICA-$\sigma$ map, and SEVEM-$\sigma$ map (dots) shown in figure~\ref{separations}, where the shaded regions correspond 
to 2 $\!\sigma$ CL.  The continuous line corresponds to the mean angular power spectra of the MC-$\sigma$ maps set.
}
 \label{powerspectrum_others}
\end{figure}

First of all, we considered the other three foreground-cleaned CMB maps to investigate the possibility that a particular pipeline in the foreground-cleaning processes 
used to produce the NILC, SMICA, and SEVEM CMB maps
could have introduced systematic effects or left residual contaminants in the CMB maps and might thus have caused a large dipole term, $\text{S}^{\texttt{Comm}}_1$, in the $\sigma$ map. 
In all these cases, we maintained the angular radius of the spherical regions as $\gamma_0 = 60^{\circ}$, and we performed the same $\sigma$ map analysis as with the \texttt{Commander} CMB map.
Each one of these three maps, NILC, SEVEM, and SMICA, and the corresponding MC CMB maps were masked using the corresponding fiducial mask. Then, the $\sigma$ maps were calculated, and their angular power spectra were obtained for the statistical significance analyses. 
The $\sigma$ maps for these three foreground-cleaned CMB maps are shown in Figure~\ref{separations}. The figure clearly shows regions in which the maximum and minimum correlations are almost antipodes in the three cases. 
In Figures~\ref{3dist-dipolos} and~\ref{powerspectrum_others} and in Table~\ref{table2}, we summarize our results regarding these tests. 
As observed, the probability for 
$\text{S}^{\texttt{MC}}_1 > \text{S}^X_1$ with $X=$ NILC, SMICA, and SEVEM is also small, from $\sim 1.4 \%$ to $\sim 2.2 \%$.  Interestingly, the direction of the maximum asymmetry, that is, the dipole direction, is almost the same for these three maps, and close to the dipole direction obtained with the \texttt{Commander} map (see the Table~\ref{table2} for a summary of our results).
We conclude that the results obtained with the NILC, SMICA, and SEVEM Planck maps are statistically consistent with the analyses done with the \texttt{Commander} Planck map.

In order to study the possibility that residual foreground or noise in the Planck CMB maps might affect our $\sigma$ map analyses, we considered them separately. 
First,  the dust and free-free 
foreground contributions\footnote{\url{https://pla.esac.esa.int}}
were used to contaminate the \texttt{Commander} and to simulate CMB MC maps. They were then masked to perform the $\sigma$ map tests. 
Furthermore, we considered the Planck noise maps 
from diverse frequencies combined with the 
\texttt{Commander} map to perform the $\sigma$ map analyses. Specifically, we used the full focal-plane Planck simulations described in detail in~\citet{Planck2015XII}, where the latest simulations set is known as {\bf FFP10}. 
These simulations contain Gaussian CMB signals and instrumental noise, and they also include, for example,  physical effects of astrophysical foreground, gravitational lensing, Doppler modulation, and frequency-dependent Rayleigh scattering effects. Additionally, this data set models the Planck mission scanning strategy, detector response, beam shape, and data reduction pipeline \citep{Pla18-ISO}. 
The results of our analyses with these data are summarized in Table~\ref{table2}.

Finally, we considered different sizes for the spherical caps, namely $\gamma_0 = 45^{\circ},\, 70^{\circ}$, and $90^{\circ}$, and repeated our directional analysis. 
Our results are also summarized in Table~\ref{table2}. 
After performing these robustness tests, 
we confirm that our results showing a dipolar asymmetry, or equivalently, a preferred axis, in the Planck CMB sphere are quite robust and have a confidence level of $98 \% -99 \%$.


To summarize, 
the analyses of individual and combined frequency CMB maps performed here and for the past 20 years strongly suggest that the north-south asymmetry phenomenon is not caused by Galactic foreground, known systematics, or instrumental noise. It instead appears to be cosmological in origin~\citep{Pla18-ISO}.

%

\section{Discussions and final remarks}\label{sec5}

In the past 20 years, many works have reported intriguing features from analyzing the large-angular properties of the CMB temperature-fluctuations maps from WMAP and Planck probes. These 
phenomena were generically termed  CMB anomalies. 
Collectively, all these analyses suggest statistically that the universe is not isotropic at large scales. 
In particular, the north-south asymmetry phenomenon robustly reveals
an axis of asymmetry on the CMB sphere that is due to anisotropic angular correlations in the Planck foreground-cleaned CMB maps. Because these results are important for the concordance model of cosmology, many analyses of the available astrophysical data are welcome. 

We investigated the angular distribution of the CMB temperature fluctuations to probe the statistical isotropy of the universe. 
To do this, we studied the TT angular correlations through a directional analysis that we applied to the four Planck foreground-cleaned CMB maps. 
Specifically, our results for the \texttt{Commander} CMB map show that the TT angular correlations exhibit 
a net dipolar feature that has the highest intensity in the direction 
$(l,b) \simeq (260^{\circ}, 130^{\circ})$
at more than
99\% confidence level (see Table~\ref{table2}). 
This highly significant dipole term observed in the $\sigma$ map
reveals the hemispherical asymmetry in the angular correlations of the CMB sky. For this reason, this phenomenon is called the  CMB north-south asymmetry. 
This result weakly depends on the size of the spherical caps that are used to scan the CMB sphere, as can be observed in Table~\ref{table2}. 
Most interestingly, the direction of the dipole is amazingly robust. It remains almost the same for the four foreground-cleaned maps, for different caps of radius $\gamma_0$, and when foreground and noise are added to the \texttt{Commander} CMB map and to the MC simulated maps. This reinforces the hypothesis that the dipolar asymmetry found in our analyses has a cosmological origin, 
that is, it appears in all foreground-cleaned maps at large scales ($\gamma_0 > 45^{\circ}$
), it is uncorrelated with instrumental noise and known Galactic and residual foregrounds, and it passed several robustness tests. 
Furthermore, the anisotropic behavior of the angular correlations we found reveals that a large part of the sky has small large-angle correlations. This is probably related to another CMB anomaly: the lack of power of the
CMB large-angle correlations \citep{Copi2006}. 
Another interesting consequence of the fact that the asymmetry we found 
reveals a preferred axis in three-space is that it might be related to the CMB anomaly called the quadrupole-octopole alignment. 
We discuss this subject in more detail in the Appendix.

More generally, several theoretical  possibilities have been proposed separately to explain some  CMB anomalies, in particular, the possible origin of the north-south asymmetry. 
This phenomenon reveals a preferred axis in three-space, the most common candidates reported in the literature are some systematic effects, such as the inhomogeneous scan of the CMB sky around the ecliptic axis, which defines a preferred axis, the mask applied to remove Galactic foregrounds, residual Galactic or extragalactic foregrounds, or a residual CMB dipole. 
Other studies 
considered modifications of the standard cosmological scenario, such as an axisymmetric primordial power spectrum, possible anisotropic inflation, primordial magnetic fields, or the topology of the universe (see, e.g., \citet{Hipolito2005,Inoue2006,B-HR,Schwarz2016,
Pereira07,Planck16-geo} and references therein). 
Nevertheless, a better understanding about the 
possible relation between (some)  CMB anomalies seems necessary for a final answer.

\begin{acknowledgements}
We acknowledge the use of data from the Planck/ESA mission, downloaded 
from the Planck Legacy Archive. 
CEK and WSHR thank  FAPES (PRONEM No 503/2020) for the financial support 
under which this work was carried out. 
AB acknowledges a CNPq fellowship. 
Some of the results in this work have been derived using the HEALPix/healpy package \footnote{\url{ http://www.eso.org/science/healpix/}} \citep{Gorski1998}.
\end{acknowledgements}

\bibliographystyle{aa}
\bibliography{references} 

\begin{appendix}
\section{Appendix}\label{appendix}
We study in this appendix the (possible) correlation 
between the statistically significant dipolar behavior of the $\sigma$ map with the quadrupole--octopole (QO) alignment of the corresponding MC CMB map (from which we obtained the $\sigma$ map). 
In other words, we investigated the possible  cause-effect phenomenon that can explain the anomalous CMB north-south asymmetry that was revealed in the $\sigma$ map analysis. 

Specifically, we wish to know whether the QO alignment is sufficient to produce $\sigma$ maps with the largest dipoles $\text{S}^{\text{MC}}_1$. 
To do this, we analyzed the quadrupoles and octopoles in detail, that is, $\ell=2, 3$, multipole components from 
the set of 50 CMB maps that produced the largest dipoles of the MC-$\sigma$ maps
$\{ \text{S}^{\text{MC}/2\sigma}_1 \}$, that is, the 
2 $\!\sigma$ CL of the set of $1000$ dipole moments $\{ \text{S}^{\text{MC}}_1 \}$. 

First, we defined a planar mode in a multipole component. 
Given a multipole $\ell$, the set of pairs of numbers $(\ell,m)$ is the mode. 
In a statistically isotropic CMB map, the power of the $\ell$ multipole, 
given by 
${\cal C}_{\ell} = (1/2\ell+1) \sum_{m=-\ell}^{\ell} \,|a_{\ell m}|^{\,2}$, 
is uniformly distributed among all the modes. 
A planar multipole means that the power of the multipole moment 
${\cal C}_{\ell}$ is concentrated in the $m = \pm \ell$ mode. 
A quadrupole, $\ell = 2$, is always planar (see, e.g.,~\cite{Abramo2009,Schwarz2016}). 

\begin{figure*}
\includegraphics[width=0.5\textwidth]{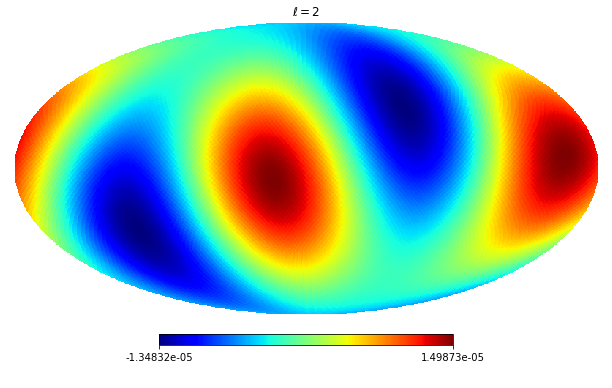} 
\includegraphics[width=0.5\textwidth]{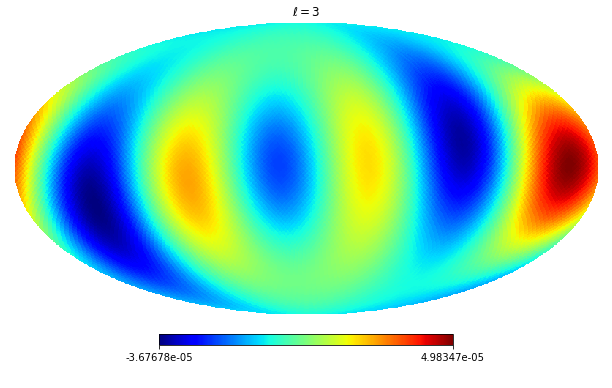} \\
\includegraphics[width=0.5\textwidth]{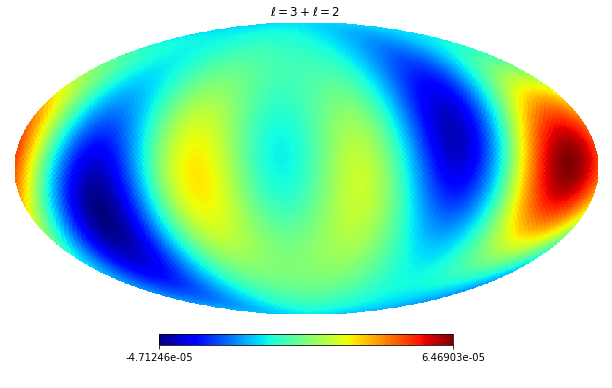}
\includegraphics[width=0.5\textwidth]{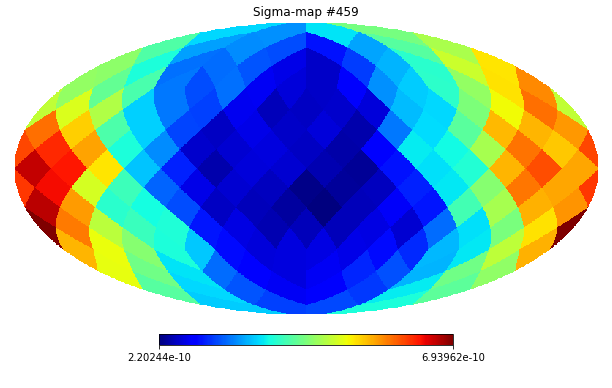}
\caption{
High $\sigma$ map dipole obtained by planarity of the octopole plus the QO alignment of an MC CMB map, part of the set 
$\{ \text{S}^{\text{MC}/2\sigma}_1 \}$. 
We show its quadrupole $\ell = 2$ (left panel, first row) and octopole $\ell = 3$ components (right panel, first row), the sum of these components (left panel, second row), and the corresponding $\sigma$ map (right panel, second row), which is obtained with all the CMB multipole components. It is clear that the sum quadrupole+octopole is dominant. 
The region with high-intensity red-blue spots clearly corresponds to the region in which the $\sigma$ map shows a positive dipole direction. 
}
\label{QO-align}
\end{figure*}

\begin{figure*}
\includegraphics[width=0.5\textwidth]{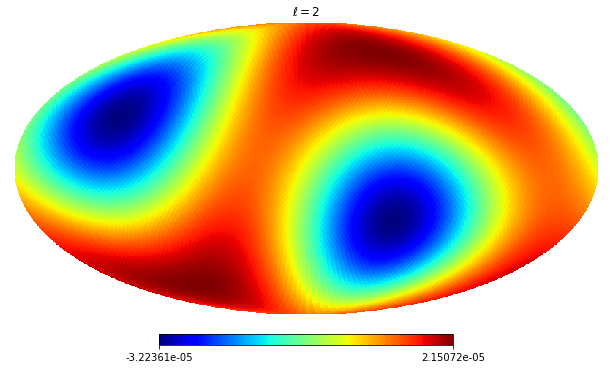} 
\includegraphics[width=0.5\textwidth]{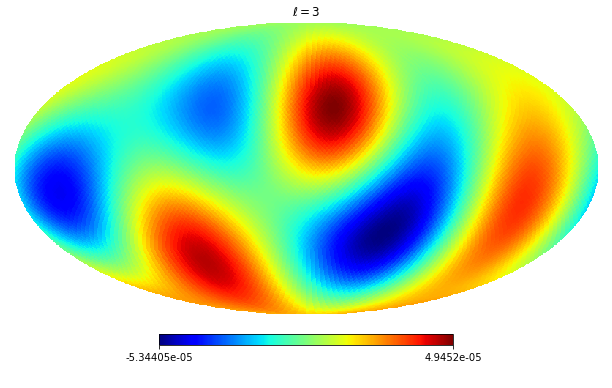} \\
\includegraphics[width=0.5\textwidth]{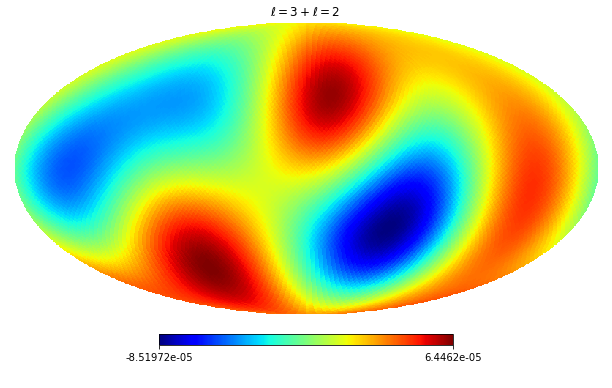}
\includegraphics[width=0.5\textwidth]{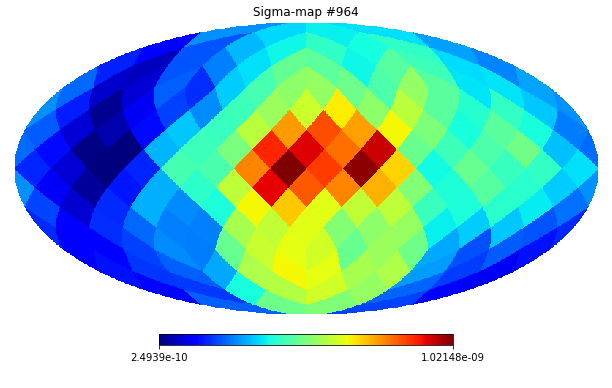}
\caption{
High $\sigma$ map dipole obtained by a coincidence of spots on the celestial sphere. 
We considered an MC CMB map, which belongs to the set $\{ \text{S}^{\text{MC}/2\sigma}_1 \}$. 
We show its quadrupole $\ell = 2$ (left panel, first row) and octopole $\ell = 3$ components (right panel, first row), the sum of these components (left panel, second row), and the corresponding $\sigma$ map (right panel, second row), which is obtained with all the CMB multipole components. The sum quadrupole+octopole is clearly dominant. 
The region with high-intensity red-blue spots corresponds to the region in which the $\sigma$ map shows a positive dipole direction. 
}
\label{manchas-coinciden}
\end{figure*}

The analysis of the full set of $1000$ MC CMB maps produced under the 
hypothesis of SI showed that the powers of the quadrupole and the octopole moments 
are, as expected\footnote{see Table 1 in~\cite{B-HR}}, uniformly distributed in their 
five modes (i.e., $(2,0), (2,\pm 1), (2,\pm 2)$) and 
seven modes (i.e., $(3,0), (3,\pm 1), (3,\pm 2), (3,\pm 3)$), respectively. 
Clearly, this uniform preference, or equal probability, also applies to the set of 50 MC CMB maps that produced the set 
$\{ \text{S}^{\text{MC}/2\sigma}_1 \}$. 
The questions is how many octopoles are planar and simultaneously have preferred planes parallel to the quadrupole plane. 
Because these events are not correlated, the probability for the two 
events to occur simultaneously is the product of the individual probabilities. 
According to the values in Table 1 in~\cite{B-HR}, 
we expect four to five cases\footnote{(the resulting probability is not an integer number)}. 
Our search for maps with these features, that is, planar octopoles plus the QO alignment, finds four MC CMB maps. 
In Figure~\ref{QO-align} we illustrate one of these interesting cases. 
For the sake of completeness, we also illustrate in 
Figure~\ref{manchas-coinciden} the case in which a high $\sigma$ map dipole can be obtained by 
coincidence\footnote{
by  coincidence we mean that the sky patch position of a red (blue) spot in $\ell=2$ is close or equal to a red (blue) spot in $\ell=3$, 
therefore, the sum of these maps results in redder (bluer) spots}
in the angular positions of the red-blue spots. 

These analyses lead us to the conclusion that the QO alignment together with the planarity of the octopole moment of the CMB map increases the probability but does not determine the north-south asymmetry phenomenon. 
The next question is what causes the simultaneous octopole planarity and the QO alignment. 
We know that both events together have a low probability of 
occurring in SI CMB maps.

In the past 20 years, many hypotheses have been investigated, but the origin of the CMB anomalies is not yet known. 
A good candidate hypothesis to explain the CMB anomalies might be spatial topology with one preferred axis: a chimney space, or a direct or $\pi$-twisted~\citep{Hipolito2005,BNPS2018}. 
A single-origin explanation for all 
CMB anomalies is still lacking so far.


\end{appendix}








   
  



\end{document}